\documentclass[a4paper,
               biblatex,     % biblatex is used
               ]{jacow}
\usepackage{pdfpages,multirow,ragged2e} %
\usepackage{graphicx,subcaption}
\makeatletter%
	\ifboolexpr{bool{xetex}}
	 {\renewcommand{\Gin@extensions}{.pdf,%
	                    .png,.jpg,.bmp,.pict,.tif,.psd,.mac,.sga,.tga,.gif,%
	                    .eps,.ps,%
	                    }}{}
\makeatother

\ifboolexpr{bool{xetex} or bool{luatex}} % test for XeTeX/LuaTeX
 {}                                      % input encoding is utf8 by default
 {\usepackage[utf8]{inputenc}}           % switch to utf8

\usepackage[USenglish]{babel}

\ifboolexpr{bool{jacowbiblatex}}%
 {%
  \addbibresource{MOP002.bib}
  \addbibresource{biblatex-examples.bib}
 }{}
\usepackage{fancyhdr}

\fancypagestyle{firstpage}{
  \fancyhf{} % clear header/footer
  \fancyfoot[C]{\small North American Particle Accelerator Conference (NAPAC2025), Sacramento, CA, USA}
  
}

\begin{document}

\title{Advancing Accelerator Virtual Beam Diagnostics through Latent Evolution Modeling: An Integrated Solution to Forward, Inverse, Tuning, and UQ Problems \thanks{Work supported by the LANL LDRD Program Directed Research (DR) project 20220074DR}}

\author{M. Rautela \thanks{mrautela@lanl.gov}, A. Scheinker, Los Alamos National Laboratory, Los Alamos, NM, US}
	
\maketitle

\maketitle
\thispagestyle{firstpage}

\abstract{
Virtual beam diagnostics relies on computationally intensive beam dynamics simulations where high-dimensional charged particle beams evolve through the accelerator. We propose Latent Evolution Model (LEM), a hybrid machine learning framework with an autoencoder that projects high-dimensional phase spaces into lower-dimensional representations, coupled with transformers to learn temporal dynamics in the latent space. This approach provides a common foundational framework addressing multiple interconnected challenges in beam diagnostics. For \textit{forward modeling}, a Conditional Variational Autoencoder (CVAE) encodes 15 unique projections of the 6D phase space into a latent representation, while a transformer predicts downstream latent states from upstream inputs. For \textit{inverse problems}, we address two distinct challenges: (a) predicting upstream phase spaces from downstream observations by utilizing the same CVAE architecture with transformers trained on reversed temporal sequences along with aleatoric uncertainty quantification, and (b) estimating RF settings from the latent space of the trained LEM using a dedicated dense neural network that maps latent representations to RF parameters. For \textit{tuning problems}, we leverage the trained LEM and RF estimator within a Bayesian optimization framework to determine optimal RF settings that minimize beam loss. This paper summarizes our recent efforts and demonstrates how this unified approach effectively addresses these traditionally separate challenges.

\maketitle

\section{Introduction}
Particle accelerators are high-dimensional spatiotemporal dynamical systems governed by hundreds to thousands of radio-frequency (RF) cavity and magnet parameters. The behavior of charged particle beams, in the form of a 6D phase space $(x, y, z, p_x, p_y, p_z)$, evolves as the beam propagates through the accelerator components. Accurate prediction and control of the beam's phase space distribution are critical for minimizing beam loss and optimizing performance. During operation, time-consuming manual tuning is often required to achieve optimal performance. With limited experimental observations of the beam, the role of virtual diagnostics becomes crucial. In recent years, machine learning models have emerged as a promising tool for non-invasive virtual beam diagnostics in particle accelerators \cite{boehnlein2022colloquium,edelen2024machine}. From a physics perspective, charged particle beam dynamics constitute a spatiotemporal phenomenon. The accelerator is a complex spatiotemporal dynamical system governed by numerous parameters. The problem of machine learning--based virtual beam diagnostics can be decomposed into multiple subproblems to systematically manage the complexity. Much of the recent progress in this direction employs standalone machine learning models to solve forward, inverse, and optimization problems.

Various ML-based models for accelerator diagnostics include integrated model-independent feedback with neural networks for automated shaping of the longitudinal phase space (LPS) of electron beams \cite{scheinker2018demonstration}, predicting LPS distributions \cite{emma2018machine,zhu2021high}, phase advance scan evaluations \cite{mayet2022predicting}, and arrival time or beam energy \cite{cropp2023virtual}. Physics-informed architectures, such as polynomial neural networks with symplectic regularization \cite{ivanov2020physics}, deep Lie map networks for magnetic field error identification \cite{caliari2023identification}, and convolutional autoencoders for relating high-dimensional phase space data to beam loss \cite{tran2022predicting}, have further enhanced predictive accuracy. Fault analysis methods include explainable SHAP-based breakdown prediction in high-gradient cavities \cite{obermair2022explainable}, classification of superconducting RF cavity faults \cite{tennant2020superconducting}, and conditional generative models for predicting errant pulses \cite{rajput2024robust}. Backcasting efforts map downstream diagnostics to upstream phase space using adaptive ML \cite{scheinker2021adaptive_SciRep}, invertible neural networks for bidirectional prediction \cite{bellotti2021fast}, tomography-based reconstruction \cite{wolski2022transverse}, and differentiable particle tracking \cite{roussel2023phase}. 

Bayesian optimization (BO) and its variants \cite{duris2020bayesian,kirschner2022tuning,breckwoldt2023machine,jalas2023tuning,ji2024multi}, deep reinforcement learning \cite{kain2020sample,meier2022optimizing,kaiser2022learning}, enhanced genetic algorithms (GAs) and multi-objective GAs (MOGAs) \cite{li2018genetic,wan2019improvement,edelen2020machine,wan2020neural} are deployed for optimization, tuning and control of accelerators. Other methods like model-independent feedback \cite{scheinker2020online}, safe extremum-seeking \cite{williams2024safe} are also utilized for tuning of accelerators. 

In this paper, we present the utility of latent evolution models (LEMs) as a unified framework that addresses three key problems in accelerator beam diagnostics: (1) forward modeling, i.e. predicting the beam's downstream phase space given upstream measurements \cite{rautela_ipac2024_mops75,rautela2024conditional}; (2a) inverse phase space reconstruction, i.e. inferring upstream beam distributions from downstream diagnostics along with aleatoric uncertainty quantification \cite{rautela_linac2024_mopb090,rautela2025time}; (2b) inverse system parameter estimation, i.e. determining the accelerator settings (such as radio-frequency cavity phases and amplitudes) given beam distribution \cite{rautela_ipac2024_mops74}; and (3) accelerator tuning/optimization, i.e. finding the optimal machine parameter settings to achieve desired beam performance (e.g. minimizing beam loss) \cite{rautela2024cbol}. Contrary to existing ML frameworks, the modular architectural design of LEMs solve multiple virtual beam diagnostics–related problems under a unified umbrella.

In the next section, we will describe the details about the latent evolution models for solving the above-mentioned problems. 

\section{Background, Formulation and Results}
Traditional machine learning approaches to dynamical systems often struggle with the high dimensionality of state representations when learning spatiotemporal dynamics. Latent evolution models overcome these challenges by factorizing the learning task into two parts: first, a dimensionality reduction stage learns an efficient spatial encoding of the system state; second, a temporal prediction stage models the evolution of that encoded state~\cite{rautela2024conditional}. The model learns spatial and temporal correlations independently through a two-step self-supervised process, which are then combined in the inference stage. LEMs feature a modular design that offers better computational efficiency, interpretability through visualization of the lower-dimensional latent space, aleatoric uncertainty quantification, multi-order autoregressive training, and improved speed for inverse and optimization problems~\cite{rautela_ipac2024_mops75,rautela2024conditional,rautela_linac2024_mopb090,rautela2025time,rautela_ipac2024_mops74,rautela2024cbol}. In the next subsections, we discuss the details of the design and working principles of LEMs for different problems of interest in virtual beam diagnostics.

\subsection{Forward Modeling of Beam Dynamics}
We define the forward problem to either predict phase space of beam from RF settings in a supervised manner or from itself in a self-supervised setting. We solve the latter problem. The forward discretized spatiotemporal beam dynamics can be written as $X_{t} = H(X_1, X_2,\dots, X_{t-1})$, where $H$ is an unknown nonlinear function, $X$ is the high-dimensional state of the system at time $t$. This can be learned as a joint probability distribution of all the states as, $P(X_1,X_2,...,X_t,...,X_T)$. The distribution can be factorised using the chain rule of probability, as $P(X_t|X_1,X_2,..X_t)$. However, in particle accelerators, $X$ is a high-dimensional object representing 6D phase space, comprising the positions and momenta of billions of particles. Using variational autoencoders (VAEs), the higher-dimensional distribution can be projected into a lower-dimensional distribution $P(\mathbf{z}_1, \mathbf{z}_2, ..., \mathbf{z}_{T})$ using Bayes' rule $p(\mathbf{z}|\mathbf{X}) \propto p(\mathbf{X}|\mathbf{z})p(\mathbf{z})$. The spatial dynamics is learned by minimizing the Evidence Lower BOund (ELBO) loss \cite{rautela2024conditional}.

After learning an appropriate latent representation, the temporal evolution in the latent space is learned with an autoregressive sequence model. Rather than modeling $X_{t+1}$ directly from $X_t$, we model the distribution $P(z_{0:T})$ of the latent trajectory $\{z_0, z_1, \dots, z_T\}$ over $T$ time steps (or modules). We factorize this joint distribution as
\begin{equation}
P(z_{0:T}) = P(z_0) \prod_{t=1}^{T} P(z_t \mid z_{t-1})~,
\end{equation}
analogous to the factorization $P(X_{0:T}) = \prod_{t=1}^{T} P(X_t \mid X_{t-1})$ in the original space. In our latent evolution framework, we do not assume the Markov property (AR(1)) for sequence modeling, which $p(z_t|z_{<t}) = p(z_t|z_{t-1})$. Instead, we utilize the full chain rule for sequence modeling i.e., AR(1:T) where the $N_{splits} \propto O(T^2)$. An LSTM or a self-attention–based transformer model can be used to capture long-term dependencies. The sequence model is not only trained to predict future latent state from a single past state, but also from all past states in AR(1:T) settings. Once the temporal model learns the dynamics, the decoder can map back the latent states to the image space.

\subsection{Inverse Problem (I): Upstream Phase Space Prediction and Aleatoric UQ}
The first inverse problem we consider is essentially the time-reversed version of the forward task: given one or more observations of the beam at downstream locations, infer its phase space at some earlier upstream location (for example, reconstruct the initial 6D phase space at the accelerator entrance). This problem is ill-posed in general (multiple upstream conditions could lead to similar downstream observations), but by learning the beam's dynamics in latent space, we can develop a robust inversion method. We implement an inverse latent evolution model, sometimes termed a reverse LEM (RLEM)\cite{rautela2025time}. Concretely, if the forward LEM learned $z_{t+1} = F_{\rm trans}(z_t)$ on sequences $\{z_0 \to z_1 \to \dots \to z_T\}$, then the RLEM is trained on sequences $\{z_T \to z_{T-1} \to \dots \to z_0\}$ so that it learns an inverse mapping. We exploit the decoupled nature of spatial and temporal learning in this problem, training only the temporal learner while keeping the spatial learner fixed.

One advantage of using a VAE-based latent space is that it naturally captures aleatoric uncertainty in the beam state. The encoder $q_\phi(z|X)$ produces a distribution (mean and variance) in latent space for a given measurement $X$. The RLEM can propagate this uncertainty backward through the transformer, yielding uncertainty estimates on the reconstructed upstream states. In other words, the probabilistic latent representation provides a way to quantify confidence in the inverse predictions: one can sample multiple $z_T$ from the encoder posterior and run the inverse model to generate an ensemble of upstream solutions, thereby obtaining error bars for $\hat{X}_0$. This uncertainty-aware inversion is a key benefit of our approach \cite{rautela2025time}.

\subsection{Inverse Problem (II): Accelerator Parameter Estimation}
The second type of inverse problem addresses the task of inferring the accelerator control parameters from observations of the beam. Specifically, given a measured 6D phase space (or its projections) at some location, we want to estimate the values of key machine settings (e.g., RF cavity phase and amplitude set-points) that produced that beam distribution \cite{rautela_ipac2024_mops74}. This is essentially a regression problem: find the function $y = g(X)$ mapping the state space to the parameter space. Directly learning a mapping from high-dimensional $X$ to $y$ is difficult due to the complexity of $X$. Our latent framework significantly simplifies this task by factorizing it into two steps: $X \to z \to y$. In other words, we treat the dependency between the state and the machine parameters as a directed acyclic graph $X \to y$; introducing the latent variable $z$ in between yields $X \to z \to y$.

Since the encoder already provides $z = E(X)$ as a compact summary of the beam, we only need to learn the mapping from $z$ to the parameter vector $y$. We employ a dense neural network (DNN) as a regressor $g_\psi: z \mapsto \hat{y}$ for this purpose. The regressor is trained on a labeled dataset of simulation runs where the accelerator settings $y$ were varied and the corresponding beam projections $X$ (and hence $z$) were recorded. Training involves minimizing the error between predicted and true parameters, for example using a mean squared error loss $\mathcal{L}_{\rm reg} = |\hat{y}(z) - y|^2$ averaged over the training set.

By utilizing the latent representation, the variational autoencoded latent regression (VALeR) model avoids having to learn the intricacies of 6D phase space directly \cite{rautela_ipac2024_mops74}. The VAE concentrates information relevant to the beam's distribution into the latent vector $z$, in effect performing feature extraction, and the DNN then operates in this compressed space where the relationships to $y$ are easier to model. 

\subsection{Tuning Problem: Latent Space Optimization}
The accelerator tuning problem is formulated as finding the set of optimal parameters $y^*$ that optimizes a desired objective function $J(y)$. In many cases the objective is to minimize beam loss or to maximize beam current through the accelerator, which can be quantified for a given machine setting by simulation or experiment. Mathematically, one seeks $y^* = \arg\min_y J(y)$. For example, $J(y)$ could be the total fractional beam loss along the accelerator for settings $y$, which is to be minimized. This is typically a high-dimensional, nonlinear optimization problem, since $y$ may consist of tens or hundreds of parameters and the beam loss depends on the collective beam dynamics in a complex way. Traditional tuning approaches (e.g. brute force scans or manual tuning) are extremely time-consuming and often cannot guarantee a true optimum. Advanced algorithms like Bayesian optimization (BO) have been successfully applied to accelerator tuning, by intelligently sampling the parameter space to find optimum solutions using as few evaluations as possible. However, directly applying BO in the full parameter space $y$ becomes intractable as dimensionality grows and when each function evaluation (a beam physics simulation or machine experiment) is costly.

Our integrated LEM approach tackles this by conducting the search in the latent state space instead of the raw parameter space \cite{rautela2024cbol}. The key idea is that the latent vector $z$ (or sequence of latents over modules) encapsulates the beam's state, and through the learned regressor $g_\psi$ we have a mapping from latent states to machine parameters $y = g_\psi(z)$. Instead of optimizing $J(y)$ over $y \in \mathbb{R}^p$ (where $p$ could be large), we can attempt to optimize $J(g_\psi(z))$ over $z \in \mathbb{R}^d$ where $d \ll p$ (e.g. an 8-dimensional latent). Intuitively, the latent space can be interpreted as a space of feasible beam configurations and by exploring it, we indirectly explore meaningful directions in the high-dimensional parameter space. We apply Bayesian optimization in the latent space, searching for the latent point $z^*$ that yields the lowest objective $J(g_\psi(z))$.

A practical complication is that not all points in latent space decode to physically realistic beam states. The VAE is trained on a finite set of beam data, so points far from the learned manifold might correspond to unphysical distributions when decoded. To address this, we incorporate a classifier-based feasibility filter as introduced in our Classifier-Pruned Bayesian Optimizer for Latent Tuning (CBOL-Tuner) approach \cite{rautela2024cbol}. This dramatically improves the success rate of finding optimal solutions that are experimentally viable. 

The CBOL-Tuner framework integrates three components: (1) the latent evolution model (CVAE + transformer) to generate temporally-structured beam trajectories in latent space, (2) the parameter estimator $g_\psi$ to map latent states to accelerator settings, and (3) the classifier-pruned Bayesian optimizer to explore the latent space for optima. By searching in the latent space of a well-trained LEM, the tuning algorithm benefits from the reduced dimensionality and the physics-informed structure of the latent manifold, yielding faster convergence to good solutions than would be possible in the raw parameter space. In our recent results, this approach not only identified multiple distinct optimal solutions for minimizing beam loss, but also outperformed standard high-dimensional global optimization methods in accuracy and precision.

\section{Conclusions}
In summary, the latent evolution modeling framework provides a common foundation for forward prediction, inverse inference, optimization, and uncertainty quantification (UQ) problems in virtual beam diagnostics. It offers a powerful way to encode complex beam dynamics into a compact and tractable representation that can be leveraged for a variety of tasks. The results for these problems have been demonstrated in recently published works.

\ifboolexpr{bool{jacowbiblatex}}%
	{\printbibliography}%
	{%
	% "biblatex" is not used, go the "manual" way
	
	%\begin{thebibliography}{99}   % Use for  10-99  references
	
}
\end{document}